\documentclass[letterpaper]{article} 
\usepackage{aaai25}  
\usepackage{times}  
\usepackage{helvet}  
\usepackage{courier}  
\usepackage[hyphens]{url}  
\usepackage{graphicx} 
\usepackage{natbib}  
\usepackage{caption} 
\usepackage[table]{xcolor}

\usepackage{subcaption}
\usepackage{booktabs}
\usepackage{multirow}
\usepackage{amsmath}
\usepackage{amssymb}
\usepackage{array}
\usepackage{pifont}
\usepackage{algorithm}
\usepackage{algorithmic}
\definecolor{mygray}{RGB}{220,220,220}

\usepackage{newfloat}
\usepackage{listings}
\DeclareCaptionStyle{ruled}{labelfont=normalfont,labelsep=colon,strut=off} 
\floatstyle{ruled}
\newfloat{listing}{tb}{lst}{}
\floatname{listing}{Listing}
\title{LEARN: Knowledge Adaptation from Large Language Model to Recommendation for Practical Industrial Application}
\author{
    Jian Jia\textsuperscript{\rm 1}, Yipei Wang\textsuperscript{\rm 2}\thanks{Work during an internship at Kuaishou Technology.}, Yan Li\textsuperscript{\rm 1}, Honggang Chen\textsuperscript{\rm 1}, Xuehan Bai\textsuperscript{\rm 1}, Zhaocheng Liu\textsuperscript{\rm 1}, Jian Liang\textsuperscript{\rm 1}, Quan Chen\textsuperscript{\rm 1}\thanks{Corresponding author.}, Han Li\textsuperscript{\rm 1}, Peng Jiang\textsuperscript{\rm 1}, Kun Gai\textsuperscript{\rm 1}
}
\affiliations{
    \textsuperscript{\rm 1}Kuaishou Technology, Beijing, China\\
    \textsuperscript{\rm 2}Southeast University, Nanjing, China \\

    \{jiajian, liyan26, chenhonggang, baixuehan03, liuzhaocheng, liangjian03, chenquan06, lihan08, jiangpeng\}@kaishou.com;230248984@seu.edu.cn;kun.gai@qq.com
}

\usepackage{bibentry}

\begin{document}

\maketitle

\begin{abstract}
Contemporary recommendation systems predominantly rely on ID embedding to capture latent associations among users and items. 
However, this approach overlooks the wealth of semantic information embedded within textual descriptions of items, leading to suboptimal performance and poor generalizations.
Leveraging the capability of large language models to comprehend and reason about textual content presents a promising avenue for advancing recommendation systems. 
To achieve this, we propose an Llm-driven knowlEdge Adaptive RecommeNdation (LEARN) framework that synergizes open-world knowledge with collaborative knowledge. 
We address computational complexity concerns by utilizing pretrained LLMs as item encoders and freezing LLM parameters to avoid catastrophic forgetting and preserve open-world knowledge. To bridge the gap between the open-world and collaborative domains, we design a twin-tower structure supervised by the recommendation task and tailored for practical industrial application.
Through experiments on the real large-scale industrial dataset and online A/B tests, we demonstrate the efficacy of our approach in industry application. We also achieve state-of-the-art performance on six Amazon Review datasets to verify the superiority of our method.
\end{abstract}
\vspace{-4mm}
\section{Introduction}
\label{sec:intro}

Inspired by the recent remarkable capabilities and rapid development of large language models (LLMs), how to introduce the rich open-world domain knowledge and great logical reasoning ability of LLMs into recommendation systems (RS) attracts attention from academics and industry.

\begin{figure}
	\centering
	\includegraphics[width=1\linewidth ]{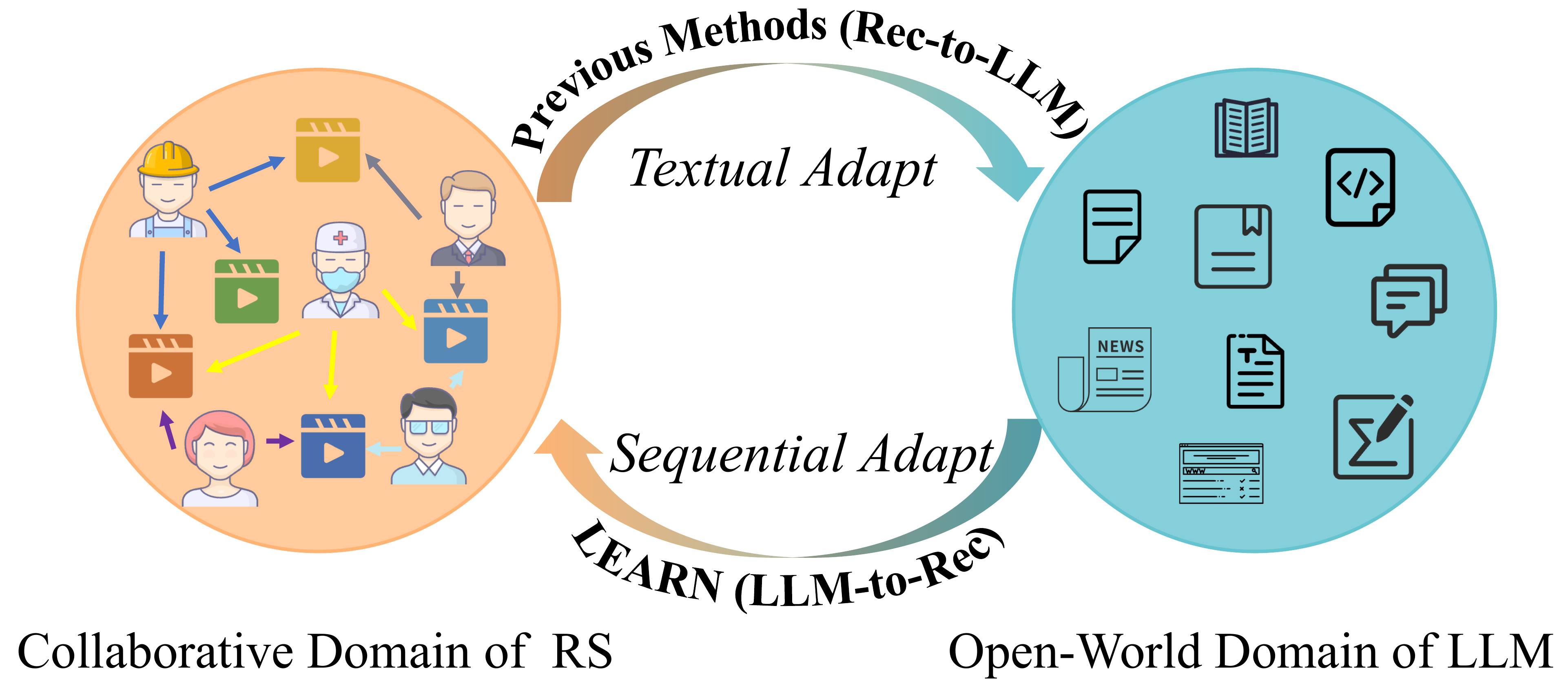}
	\caption{Illustration of ``Rec-to-LLM" and ``LLM-to-Rec" approaches. Previous methods textualize recommendation data to natural language conversations, which are then fed into LLM to obtain text predictions. In contrast, our method, LEARN, transforms item text information into LLM embedding, which is projected into the collaborative domain to achieve alignment with industry recommendation tasks.}
	\label{fig:compare}
	\vspace{-5mm}
\end{figure}

Current RS heavily rely on distinct ID embeddings and focus on capturing latent user-item associations based on historical interactions. This approach overlooks the semantic information in item text descriptions and struggles to generalize to unseen data, resulting in suboptimal performance in industrial cold-start scenarios and long-tail user recommendations. Moreover, unlike the fields of computer vision (CV) and natural language processing (NLP), the ID-embedding-based modeling approach in RS does not facilitate the development of a pretrained model that can perform well across downstream tasks and subscenarios.

To alleviate the poor generalization of current RS, various methods are proposed to utilize text information and integrate LLM with RS to generate textual predictions such as user interests \cite{ren2024representation}, next item information \cite{li2023prompt}, and recommend reasons \cite{zhang2023recommendation}. 

Previous research \cite{lyu2023llm, liu2023chatgpt, zhang2023chatgpt, gao2023chat, zs-llm4rec-sanner2023large, hou2024large} on integrating recommendation systems with LLMs typically follows a unified strategy termed the ``Rec-to-LLM" adaptation in this paper. This strategy involves adapting user-item interaction data from the recommendation domain (target domain) into the textual format of the LLM open-world domain (source domain), as depicted in Fig.~\ref{fig:compare}. Specifically, these methods \cite{liao2023llara, bao2023tallrec, yuan2023go, bao2023bi, lin2023rella} design task-specific prompts to transform recommendation data to conversational formats and employ the next token prediction loss, aligning the input organization and target tasks with those of the LLM pre-training stage.

However, our empirical investigations reveal that the ``Rec-to-LLM" adaptation fails to yield practical benefits in real-world industry applications. This inefficacy can be attributed to the inherent shortcomings of this approach. First, given the input length limitations (2K to 128K) and computation complexity of LLMs, inferring or finetuning LLM with the textualized interactions is unaffordable in industry scenarios. In our short video platform, users interact with nearly 800 short videos on average each week. Therefore, handling the global user history interactions over several months with LLM poses significant computational burdens.
Second, finetuning LLM with recommendation data often leads to catastrophic forgetting and suboptimal performance, due to the significant domain gap between collaborative knowledge of RS and general open-world knowledge of LLM. Third, the misalignment between training objectives of LLM and recommender further prevent previous work from achieving satisfactory performance.

To overcome the limitations noted, we introduce the Llm-driven knowlEdge Adaptive RecommeNdation (LEARN) approach, designed to synergize the open-world knowledge of LLMs with the collaborative knowledge of recommenders. Contrary to previous methods following ``Rec-to-LLM" adaptation, our approach adapts knowledge from LLM to recommendation (LLM-to-Rec) depicted in Fig.~\ref{fig:compare}. 
We employ the LLM as a content extractor, with the recommendation task serving as the training target. Specifically, the proposed LEARN framework consists of a user tower and an item tower.
Both towers consist of the Content EXtraction (CEX) and Preference ALignment (PAL) modules. 
To address the computational challenges associated with processing extensive user history interaction, the CEX module employs the pretrained LLM as an item encoder rather than as a user preference encoder. To avoid catastrophic forgetting of open-world knowledge, we freeze the LLM during the training stage. 
Furthermore, to bridge the domain gap between open world and collaborative knowledge, we design the PAL module and adopt the self-supervised training target of the recommendation task to guide model optimization. 
The user and item embeddings generated by LEARN are taken as inputs of the online ranking model.

To validate our methods in real industry application, we build a large-scale dataset collected from the real recommendation scenario and evaluate our method in the online A/B test. Experiments are also conducted on Amazon Reviews datasets \cite{ni2019justifying} to make a fair comparison with previous methods. State-of-the-art performance is achieved in three metrics of six datasets and verifies the superiority of the LEARN framework.

We conclude our contributions as follows:
\begin{itemize}
\item We propose the Llm-driven knowlEdge Adaptive RecommeNdation (LEARN) framework to efficiently aggregate the open-world knowledge encapsulated within LLMs into RS.
\item We propose the CEX and PAL modules to solve the catastrophic forgetting of open-world knowledge in LLM and take the recommendation task to bridge the domain gap between open-world and collaborative knowledge.  
\item We evaluate our method on the large-scale industry dataset and validate it through online A/B testing, verifying its profitability in real-world industry scenarios.
\item To confirm the superiority of our method over previous work, we conduct experiments on six public datasets and achieve SOTA performance in three metrics, particularly bringing an average 13.95\% improvement in Recall@10.
\end{itemize}

\section{Related Work}
\subsection{Content-based Recommendation.}

Traditional RSs are predominantly based on ID-based embeddings \cite{hidasi2015session, kang2018self, sun2019bert4rec, li2022recguru}, which frequently suffer from limited generalizability. To address this, extensive research has focused on deepening the understanding of user and item content to bring incremental information and enhance the generalization capabilities for online RS. 
Wu \textit{et al.} developed the large-scale MIND text dataset \cite{wu2020mind} specifically for the news recommendation task, advancing research on the impact of understanding text content for RS recommendation systems. 
Following this, various studies leverage the BERT \cite{devlin2018bert} model to improve content understanding. ZESRec \cite{ding2021zero}, UniRec \cite{hou2022towards}, and TBIN \cite{chen2023tbin} take the pretrained BERT model as the encoder to extract content embedding for item text description. Recformer \cite{li2023text} draws inspiration from BERT's training mechanism, combining masked language model loss with contrastive loss, and features a redesigned tokenizer to encode textual information of items.
In addition to utilizing textual information, some methods also attempt to incorporate visual information into the recommendation model. SimTier and MAKE \cite{sheng2024enhancing} adopt CLIP \cite{radford2021learning} and MoCo \cite{he2020momentum} to extract image features. MoRec \cite{yuan2023go} and MISSRec \cite{wang2023missrec} incorporate visual content from item images using ResNet \cite{he2016deep} and ViT \cite{dosovitskiy2020image} in the sequential recommendation.

\begin{figure*}[h]
    \centering
    \includegraphics[width=\textwidth]{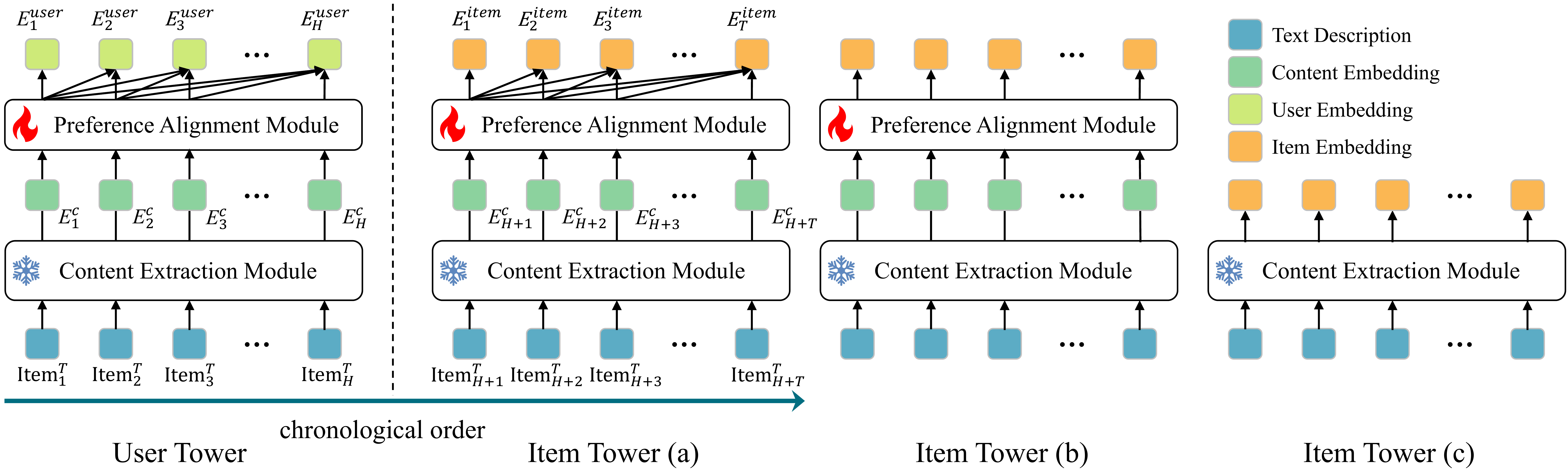}
    \caption{Illustration of our Llm-driven knowlEdge Adaptive RecommeNdation (LEARN) framework. The LEARN framework employs a twin-tower architecture comprising a user tower and an item tower. The user tower processes history interactions to generate user embeddings $E^{user}$, while the item tower handles target interactions to produce item embeddings $E^{item}$. User Tower and Item Tower (a) leverage the causal attention mechanism. Item Tower (b) adopts a self-attention mechanism.  Without the preference alignment module, Item Tower (c) directly utilizes the content embedding as the item embedding.}
    \label{fig:tow_tower}
    \vspace{-3mm}
\end{figure*}

\subsection{LLM-based Recommendation.} 
Due to the powerful capabilities demonstrated by LLM in textual understanding and common sense reasoning, an increasing number of studies are exploring the integration of LLM in recommender systems (RS) \cite{lin2023can, fan2023recommender}. The first type freezes the LLM parameters and takes the LLM as a recommender. Some works \cite{gao2023chat, liu2023chatgpt, zhang2023chatgpt, zhang2024recgpt, hou2024large, xi2023towards} proposed the task-specific prompt to construct recommendation dialogues and used ChatGPT to generate candidates. RLMRec \cite{ren2023representation} utilized ChatGPT to generate user/item profiles.
The second type fine-tunes the LLM on specially prepared textual datasets from the recommendation domain. LlamaRec \cite{yue2023llamarec} took the item title as textual data and optimized the LLM by ranking scores. TALLRec \cite{bao2023tallrec} proposed a two-stage tuning framework and finetunes the LLM with LoRA \cite{hu2021lora} for few-shot recommendation. LLaRA \cite{liao2023llara} combined the LLM prompt with ID embeddings to align the LLM and the well-established sequential recommenders. ReLLa \cite{lin2023rella} proposed a retrieval-enhanced instruction tuning method and finetuned Vicuna \cite{chiang2023vicuna} on a mixed dataset.

Both types of prior research adapt the user-item interaction data of recommendation systems to the textual conversation format of LLMs and utilize the training loss of LLMs to finetune the model. These methods transfer data and tasks from the recommendation domain (target domain) to the LLM domain (source domain), and are therefore referred to as ``Rec-to-LLM" methods in this paper.

\section{Method}

\subsection{Model Architecture}

Given user history interactions in chronological order, the interaction sequence is split into two segments based on a specific timestamp: the first segment is the history interaction sequence $U^{hist}$, and the second segment is the target sequence $U^{tar}$. The length of history and target interactions are denoted as $H$ and $T$, respectively. We propose the LEARN framework to capture the user's interests from the history interaction and predicts the next item the user is interested in. The LEARN framework consists of a user tower and an item tower as shown in Fig.~\ref{fig:tow_tower}.

\subsubsection{User Tower}

\begin{figure}[t]
	\centering
	\includegraphics[width=1\linewidth ]{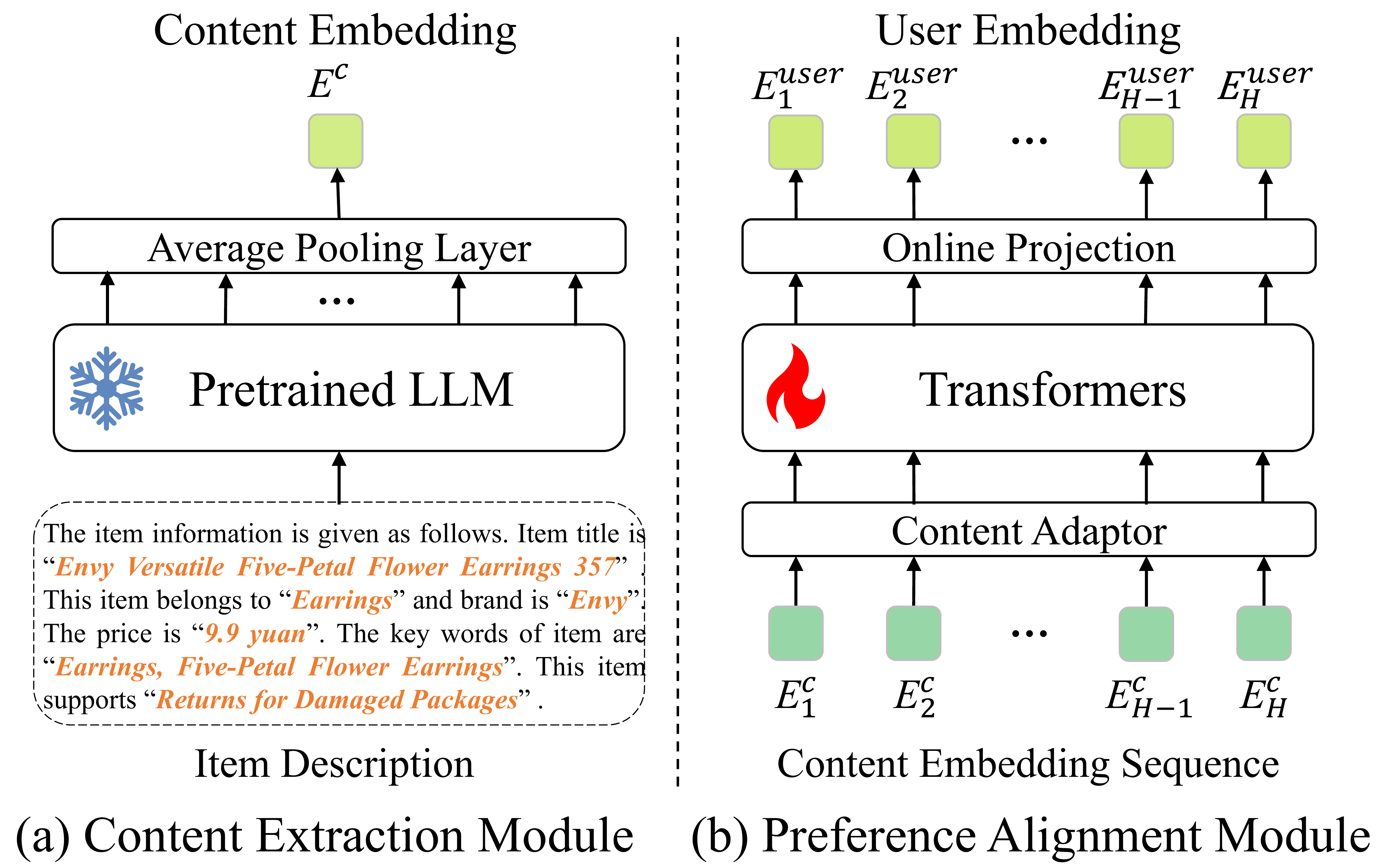}
    \vspace{-1em}
    \caption{Illustration of the Content Extraction (CEX ) module and Preference ALignment (PAL) module. The CEX module utilizes a pretrained LLM to generate content embeddings from item text descriptions. The PAL module takes these content embeddings and projects them from the open-world domain into the collaborative domain embeddings used in the online recommender.}
    \label{fig:two_module}
    \vspace{-5mm}
\end{figure}

The User Tower comprises a Content EXtraction (CEX) module and a Preference Alignment (PAL) module, as depicted in Fig.~\ref{fig:two_module}. The input of the user tower is a sequence of history items that interact with the user. 
Each item is described textually according to the prompt template shown in Fig.~\ref{fig:two_module}. 
The prompt is designed to be highly concise to effectively assess the informativeness of the textual description.
The CEX module processes these item descriptions employing a pre-trained LLM and an average pooling layer to generate content embeddings $E^{c}$. 
During training, parameters of the pretrained LLM remain frozen, and the hidden states of the final decoder layer are used as output embeddings, which are subsequently sent to the pooling layer, as illustrated in Fig.~\ref{fig:two_module}(a). 
For the entire history interaction sequence $U^{hist}$, the CEX module converts the textual description of each item into a content embedding $E^{c}$, forming a content embedding sequence. Each item is processed independently by the CEX module.

The Preference ALignment (PAL) module captures user preference and outputs user embeddings based on the content embedding sequence. The PAL module starts with a content adaptor to perform dimension transformation. Subsequently, the transformer with 12 layers serves as the backbone network, following the configuration of BERT-base \cite{devlin2018bert} model. This transformer is specifically designed to learn implicit item relationships and model user preferences. Unlike bidirectional attention in BERT \cite{devlin2018bert}, our module employs the causal attention mechanism to model sequential dependency by focusing exclusively on past items, in line with the chronological nature of user preferences. The output embeddings of our transformer are further processed through online projection layers to produce user embeddings $E^{user}\in \mathbf{R}^{64}$, which are directly utilized for online e-commerce recommender in Fig.~\ref{fig:cvr_model}.

\subsubsection{Item Tower}

The item tower processes textual descriptions of item content and outputs item embedding $E^{item}$ tailored for the recommendation domain. As depicted in Fig.~\ref{fig:tow_tower}, three variants of the item tower are proposed.

ItemTower(a) employs the same causal attention mechanism as the User Tower, whereas ItemTower(b) adopts the self-attention mechanism where each item attends exclusively to its own content.
Despite this difference, both variants maintain an identical model architecture and share the same weights with the UserTower. 
To bridge the substantial gap between open-world and collaborative domains, both variants project content embeddings from the LLM into user/item embeddings used in recommendation system, following the ``LLM-to-Rec" approach. In contrast, ItemTower(c) directly leverages the content embedding $E^{c}$ as the item embedding $E^{item}$ to guide user perference learning in the ``Rec-to-LLM" manner. In the training stage, ItemTower(a) processes the entire user target sequence $U^{tar}$ as input, while ItemTower(b) and ItemTower(c) handle individual items independently. In the inference stage, all three variants take a single item as input and generate item embedding independently. Due to superior performance in Tab.~\ref{tab:ablation_indus}, ItemTower(a) is adopted as the default setting.

\subsubsection{Training Target} \label{sec:training_target}

To bridge the gap between content embeddings in the open-world domain of LLMs and user/item embeddings in the collaborative domain of RS, we align our training objectives with those of the online ranking model.
In an online RS, the ranking model computes similarities between the user embedding and the embeddings of all items in the gallery. The top k items with the highest similarity scores are identified as those likely to interest the user. 
Thus, we employ a self-supervised contrastive learning mechanism to model user preferences, aligning with the goals of the online RS.
This approach maximizes the similarity between the user embedding and the embeddings of relevant items while minimizing similarities with irrelevant items. We sample user embeddings from the user history sequence and item embeddings from the target sequence of the same user to construct positive sample pairs. Target item embeddings of other users in the same batch are sampled as negative. To fully exploit user interactions and capture user long-term interest, we adopt the dense all action loss \cite{pancha2022pinnerformer}. $N_{h}$ user embeddings are sampled from the history sequence and the $N_{t}$ element embeddings are sampled from the target sequence. This allows us to construct $N_{h} \times N_{t}$ positive sample pairs from single user interaction data to apply dense all action loss. $N_{h}$ and $N_{t}$ are set to 10 by default.

\subsubsection{Sampling Strategy}

Although items have been sampled based on the importance of behavior during the construction of the industry dataset, the length of user sequences remains excessively long due to training resource constraints. To address this, we design a two-stage sampling strategy during the training phase. In the first stage, we perform random sampling from the complete user history/target interactions to serve as input for the User Tower, ensuring that the data used for modeling user interests is unbiased.  Visualization of user interaction data revealed that items -- users recently interacted with -- better reflect user current interests and are more relevant to the target items of user preferences. 
Thus, in the second stage, when constructing positive and negative sample pairs, we implement a sample weighting strategy that prioritizes recent items. The weighting $\tilde{w}_i$ of the $i$-th item in the history/target sequence is computed as:
\begin{align} \label{eq:weight}
\tilde{w}_i = \frac{w_i}{\max(w)}, \quad
\text{where} \  w_i = \log (\alpha + i \cdot \frac{\beta - \alpha}{N - 1}). 
\end{align}
The hyperparameters $\alpha$ and $\beta$ are set to 10 and 10000. $N$ is the length of user history/target interactions sampled from the first stage.

\section{Experiments}

\subsection{Experimental Settings}

\subsubsection{Dataset}
For industry application, we build a large-scale practical recommendation dataset from the e-commerce platform of a short video application. The industry dataset compose 12 million users interacted with 31 million items over a 10-month period from June 2022 to April 2023. The interactions from the first nine months are used as historical data, while the interactions from the last month are designated as target data. 
Six types of information (\textit{title, category, brand, price, keywords, attributes}) are collected to form the item description. To make a fair comparison, we adopt the widely used Amazon Review dataset \cite{ni2019justifying} and follow the same settings as RecFormer \cite{li2023text}. Seven categories are selected as pretraining data and the other six categories are selected as finetuning data to evaluate our method. Three types of information (\textit{title, category, brand}) are used to form the item description. The statistics of public and industry datasets are shown in Tab.~\ref{tab:stat_amazon} and Tab.~\ref{tab:stat_kuaishou}.

\begin{table}[t]
\small
\centering
\resizebox{1\columnwidth}{!}{%
\begin{tabular}{lrrrrr}
\toprule
\multicolumn{1}{c}{\textbf{Datasets}} & \multicolumn{1}{c}{\textbf{\#Users}} & \multicolumn{1}{c}{\textbf{\#Items}} & \multicolumn{1}{c}{\textbf{\#Inters.}} & \multicolumn{1}{c}{\textbf{Avg. n}} & \multicolumn{1}{c}{\textbf{Density}} \\ \midrule
\textbf{Pre-training} & 3,613,906 & 1,022,274 & 33,588,165  & 9.29& 9.1e-6\\
-Training  & 3,501,527 & 954,672   & 32,291,280  & 9.22& 9.0e-6\\
-Validation& 112,379   & 67,602& 1,296,885   & 11.54& 1.7e-4\\ \midrule
\textbf{Scientific}   & 11,041& 5,327& 76,896& 6.96& 1.3e-3\\
\textbf{Instruments}  & 27,530& 10,611& 231,312& 8.40& 7.9e-4\\
\textbf{Arts}   & 56,210& 22,855& 492,492& 8.76& 3.8e-4\\
\textbf{Office} & 101,501   & 27,932& 798,914& 7.87& 2.8e-4\\
\textbf{Games}  & 11,036& 15,402& 100,255& 9.08& 5.9e-4\\
\textbf{Pet}& 47,569& 37,970& 420,662& 8.84& 2.3e-4\\ 
\bottomrule
\end{tabular}}
\vspace{-2mm}
\caption{Statistics of the Amazon Review datasets. Avg. n denotes the average length of user interactions.}
\label{tab:stat_amazon}
\vspace{-2mm}
\end{table}

\begin{table}[t]
\small
\centering
\resizebox{1\columnwidth}{!}{%
\begin{tabular}{lrrrrr}
\toprule
\multicolumn{1}{c}{\textbf{Datasets}} & \multicolumn{1}{c}{\textbf{\#Users}} & \multicolumn{1}{c}{\textbf{\#Items}} & \multicolumn{1}{c}{\textbf{\#Inters.}} & \multicolumn{1}{c}{\textbf{Avg. n}} & \multicolumn{1}{c}{\textbf{Density}} \\ \midrule
\textbf{Train}   & 11,965,799 & 31,044,924 & 1,883,216,224 & 157.38 & 5.07e-6\\
\textbf{Test}  & 239,813& 6,519,074 & 37,683,369 & 157.13 & 2.41e-5\\
\bottomrule
\end{tabular}}
\vspace{-2mm}
\caption{Statistics of the industry datasets.}
\label{tab:stat_kuaishou}
\vspace{-3mm}
\end{table}

\begin{table*}[t]
\centering
\small
\scalebox{0.9}{
\setlength{\tabcolsep}{1mm}{
\begin{tabular}{llccccccccccc}
\toprule
\multicolumn{1}{c}{\textbf{}} & \multicolumn{1}{c}{\textbf{}} & \multicolumn{4}{c}{\textbf{ID-Only Methods}} & \multicolumn{2}{c}{\textbf{ID-Text Methods}} & \multicolumn{4}{c}{\textbf{Text-Only Methods}} & \multirow{2}{*}{\textbf{Improv.}} \\ 
\cmidrule(lr){3-6} \cmidrule(lr){7-8} \cmidrule(lr){9-12}
\multicolumn{1}{c}{\textbf{Dataset}} & \multicolumn{1}{c}{\textbf{Metric}} & \textbf{GRU4Rec} & \textbf{SASRec} & \textbf{BERT4Rec} & \textbf{RecGURU} & \textbf{FDSA} & \textbf{S$^3$-Rec} & \textbf{ZESRec} & \textbf{UniSRec} & \textbf{RecFormer} & \textbf{LEARN}  \\ \midrule
\multirow{3}{*}{Scientific} & NDCG@10 & 0.0826 & 0.0797 & 0.0790 & 0.0575& 0.0716& 0.0451& 0.0843 & 0.0862 & \underline{0.1027} & \textbf{0.1060} & + 3.21\%  \\
& Recall@10& 0.1055& 0.1305& 0.1061 & 0.0781& 0.0967& 0.0804& 0.1260& 0.1255& \underline{0.1448} & \textbf{0.1594} & +10.08\%  \\\midrule
\multirow{3}{*}{Instruments}& NDCG@10  & 0.0633& 0.0634& 0.0707 & 0.0468& 0.0731& 0.0797 & 0.0694& 0.0785& \underline{0.0830} & \textbf{0.0878} & + 5.78\%\\
& Recall@10& 0.0969& 0.0995& 0.0972 & 0.0617& 0.1006& { 0.1110}& 0.1078& \underline{0.1119}  & 0.1052  & \textbf{0.1240} & +17.87\% \\ \midrule
\multirow{3}{*}{Arts}& NDCG@10  & { 0.1075}& 0.0848& 0.0942 & 0.0525& 0.0994& 0.1026& 0.0970& 0.0894& \textbf{0.1252} & \underline{0.1225} &  -- \\
& Recall@10& 0.1317& 0.1342 & 0.1236 & 0.0742& 0.1209& 0.1399 & 0.1349 & 0.1333 & \underline{0.1614} & \textbf{0.1701} & + 5.39\% \\\midrule
\multirow{3}{*}{Office}   & NDCG@10  & 0.0761& 0.0832 & { 0.0972} & 0.0500 & 0.0922 & 0.0911& 0.0865 & 0.0919 & \underline{0.1141} & \textbf{0.1167} &  + 2.28\% \\
& Recall@10 & 0.1053 & 0.1196 & 0.1205 & 0.0647 & 0.1285 & 0.1186 & 0.1199 & 0.1262 & \underline{0.1403} & \textbf{0.1549} & +10.41\%\\\midrule
\multirow{3}{*}{Games} & NDCG@10  & 0.0586 & 0.0547 & 0.0628 & 0.0386 & 0.060 0& 0.0532 & 0.0530 & 0.0580 & \underline{0.0684} & \textbf{0.0798} & +16.67\%\\
& Recall@10 & 0.0988 & 0.0953 & 0.1029 & 0.0479& 0.0931 & 0.0879 & 0.0844 & 0.0923 & \underline{0.1039} &  \textbf{0.1345} & +29.45\%\\ \midrule
\multirow{3}{*}{Pet} & NDCG@10  & 0.0648 & 0.0569 & 0.0602 & 0.0366 & 0.0673 & 0.0742 & 0.0754 & 0.0702 & \underline{0.0972} & \textbf{0.0990} & + 1.85\%  \\
& Recall@10 & 0.0781 & 0.0881 & 0.0765 & 0.0415 & 0.0949 & 0.1039 & 0.1018 & 0.0933 & \underline{0.1162} & \textbf{0.1284} & +10.50\% \\\bottomrule
\end{tabular}
}}
\vspace{-2mm}
\caption{Performance comparison of different recommendation models. The best and second best performances are bold and underlined, respectively. Improv. denotes the relative improvement over the SOTA method RecFormer. }
\label{tab:perf_amazon}
\vspace{-5mm}
\end{table*}

\subsubsection{Implementation Details}
Baichuan2-7B \cite{baichuan2023baichuan2} is adopted as the LLM to extract content embedding based on item text description due to its robust capabilities in understanding Chinese and English text. The LLM parameters are frozen during the training stage. All experiments take the AdamW optimizer and the cosine scheduler as default settings.
For industry datasets, the training batch size is set to 240 and the length of user history and target interaction are set to 80 and 40 due to memory limitation. Training epochs are set to 10. The hit rate (H@50, H@100) and recall (R@50, R@100) at Top50 and Top100 is used for performance evaluation.
For the Amazon Review datasets, the batch size is set to 1024 during pretraining and 16 during fine-tuning. The learning rate is 5e-5 and 2e-5, respectively, for the two stages. Training epochs are set to 20 for pretraining and 200 for fine-tuning. We follow the evaluation settings proposed by RecFormer, applying the leave-one-out strategy \cite{kang2018self} for the evaluation. Three metrics — NDCG@10 (N@10), Recall@10 (R@10), and MRR — are used to ensure a fair comparison. Given the limited interaction sequence length in the Amazon Review dataset, we opt not to apply any sampling strategy in the training stage of LEARN.

\subsection{Performance on Amazon Review}
\subsubsection{Overall Performance}
\begin{figure}[ht]
	\centering
	\includegraphics[width=1\linewidth ]{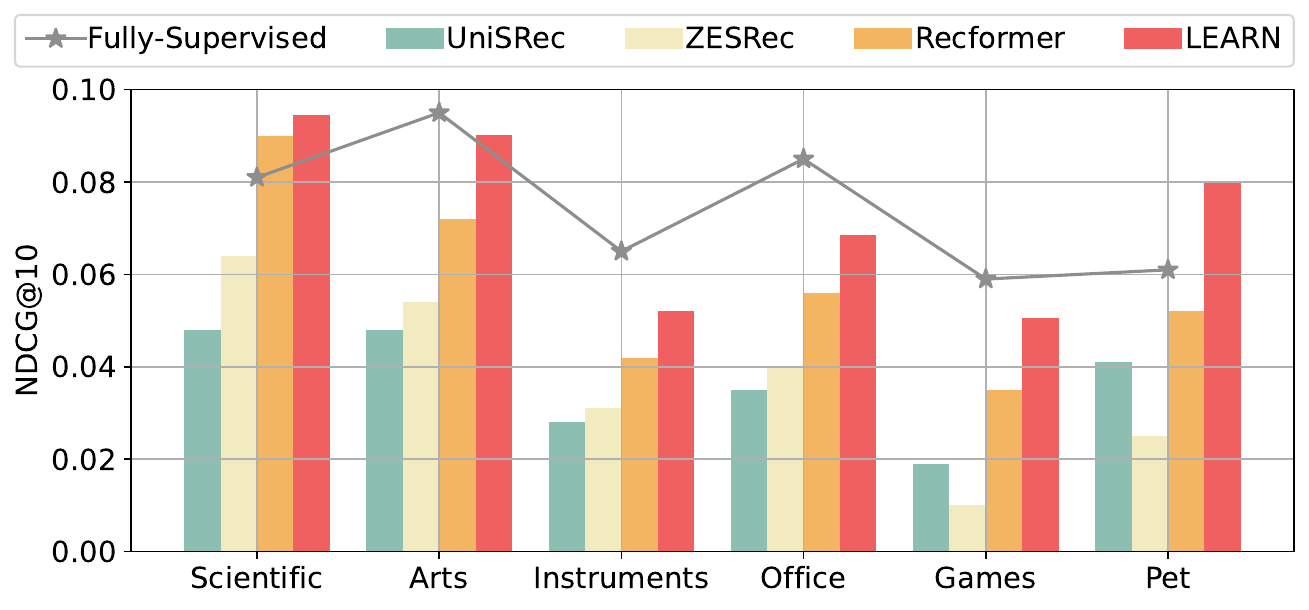}
    \vspace{-4mm}
	\caption{Performance comparison with text-only SOTA methods under zero-shot setting. }
	\label{fig:zero_shot}
    \vspace{-7mm}
\end{figure}

To verify the effectiveness of our method, performance on Amazon Review is reported in Tab.~\ref{tab:perf_amazon}. 
We compare LEARN with three categories of methods: ID-Only methods (GRU4Rec \cite{hidasi2015session}, SASRec \cite{kang2018self}, BERT4Rec \cite{sun2019bert4rec}, RecGURU \cite{li2022recguru}) , ID-Text methods (FDSA \cite{zhang2019feature}, S$^3$-Rec \cite{zhou2020s3}), and Text-only methods (ZESRec \cite{ding2021zero}, UniSRec \cite{hou2022towards}, RecFormer \cite{li2023text}). Our method achieves significant improvements compared to the ID-only, ID-Text, and Text-only methods. Specifically, compared to the SOTA method RecFormer, our method LEARN brings improvements of 10.08\%, 17.87\%, 5.39\%, 10.41\%, 29.45\%, and 10.50\% in Recall@10 on the Scientific, Instruments, Arts, Office, Games, and Pet datasets, respectively. Instead of using masked language modeling (MLM) loss and a two-stage fine-tuning process like RecFormer, the proposed LEARN model is supervised solely by user-item contrastive loss in a single fine-tuning stage. Our method achieves significant performance improvements despite fewer loss constraints and a simpler training process, further demonstrating the effectiveness of our framework. We also conduct experiments in zero-shot settings (only with the pre-training stage) following the RecFormer setting. Results in Fig.~\ref{fig:zero_shot} demonstrate that our LEARN framework can be taken as a pretrained recommendation model and performs well in downstream subscenarios. The performance of other methods in Fig.~\ref{fig:zero_shot} is referenced in RecFormer.

\subsubsection{Ablation Study}
The insight behind our performance improvement is that: a significant gap between the collaborative domain of the recommenders and the open-world domain of LLMs. Given the informative content embedding as inputs, one efficient way to bridge this gap is taking user-item interactions as alignment target to transform content embeddings into user/item embeddings. We demonstrate our insight in Tab.~\ref{tab:ablation_amazon}. 
First, we calculate the user embedding by averaging the content embeddings of all items the user has interacted with and use content embeddings of items as item embeddings directly. Since there is no alignment between the LLM and recommendation domains, we term this method ``w/o Align". We find that the performance of the ``w/o Align" is very poor, which confirms our hypothesis that there is a significant gap between the open-world knowledge of the LLM domain and the collaborative knowledge of the recommendation domain. Therefore, content embeddings generated by LLMs are not suitable for direct use in recommendation tasks.
Second, we use the content embedding generated by the LLM as the alignment target, with ItemTower(c) shown in Fig.~\ref{fig:tow_tower}. ItemTower(c) variant achieves domain alignment by transforming the recommendation domain into the LLM domain, following the same ``Rec-to-LLM" adaptation of previous work. Experiments reveal that ItemTower(c) variant is inferior to LEARN. This is because the distinct characteristics of recommendation knowledge (target domain) are not well represented in the LLM’s open-world knowledge (source domain). In contrast, by projecting the source domain into the target domain space, the LEARN model is better attuned to the complex and intricacies of the recommendation (target) distribution, leading to improved performance.

It is worth noting that for the Amazon Review dataset, ItemTower(a) is equivalent to ItemTower(b), as the length of the user target sequence is 1 due to the leave-one-out setting \cite{kang2018self, li2023text}.

\begin{table}[t]
\centering
\small
\scalebox{0.95}{
\setlength{\tabcolsep}{1mm}{
\begin{tabular}{llcccc}
\toprule
\textbf{Dataset} & \textbf{Metric} &  \textbf{w/o Align} & \textbf{w/ ItemTower(c)}  & \textbf{LEARN}   \\ \midrule
\multirow{2}{*}{Scientific}& N@10& 0.0504 & 0.0978 & 0.1060 (+ 8.38\%) \\
& R@10& 0.0813 & 0.1389  & 0.1594 (+14.76\%) \\ \midrule
\multirow{2}{*}{Instruments}& N@10& 0.0164 & 0.0679 & 0.0878 (+29.31\%) \\
& R@10& 0.0332 & 0.0940 & 0.1240 (+31.91\%)  \\ \midrule
\multirow{2}{*}{Arts}& N@10 & 0.0308 & 0.0900 & 0.1225 (+36.11\%) \\
& R@10& 0.0633 & 0.1337 & 0.1701 (+27.23\%) \\ \midrule
\multirow{2}{*}{Office}& N@10& 0.0166 & 0.0890 & 0.1167 (+31.12\%) \\
& R@10& 0.0312 & 0.1144 & 0.1549 (+35.40\%) \\ \midrule
\multirow{2}{*}{Games}& N@10& 0.0175 & 0.0555& 0.0798 (+43.78\%)  \\
& R@10& 0.0361 & 0.0891 & 0.1345 (+50.95\%) \\ \midrule
\multirow{2}{*}{Pet} & N@10 & 0.0312 & 0.0812 & 0.0990 (+21.92\%)  \\
& R@10& 0.0427 & 0.1010& 0.1284 (+27.13\%)\\ \bottomrule
\end{tabular}}}
\vspace{-2mm}
\caption{Ablation studies of alignment strategy on Amazon Review. \textit{w/o Align} averages the content embeddings of all interacted items to create the user embedding. \textit{w/ ItemTower(c)}  take the content embedding of items to supervise the user embedding learning.}
\label{tab:ablation_amazon}
\vspace{-3mm}
\end{table}

\subsection{Performance on Industry dataset}

\begin{table}[t]
\centering
\resizebox{\columnwidth}{!}{%
\begin{tabular}{l|cccc}
\hline
 \textbf{Ablation} & \textbf{H@50} & \textbf{R@50} & \textbf{H@100} & \textbf{R@100} \\ \hline
 w/o Align &  0.0069 & 0.0154 & 0.0101 & 0.0210 \\
 w/ ItemTower(c) & 0.0292 & 0.0416 & 0.0468 & 0.0626 \\
 w/ ItemTower(b) & 0.0313 & 0.0488 & 0.0505 & 0.0675 \\
 w/ RandomSample & 0.0440 & 0.0610 & 0.0701 & 0.0905  \\
LEARN (ours)  & \textbf{0.0477} & \textbf{0.0663} & \textbf{0.0751} & \textbf{0.0970}  \\ \hline
\end{tabular}%
}
\vspace{-2mm}
\caption{Ablation studies of alignment strategy and sampling strategy on industry dataset.}
\label{tab:ablation_indus}
\vspace{-3mm}
\end{table}

\subsubsection{Ablation Study} To further validate the rationality of our model design, we conduct ablation studies on the large-scale dataset collected from the real industry scenario. Dataset details are given in Tab.~\ref{tab:stat_kuaishou}. 
As shown in Tab~\ref{tab:ablation_indus}, ``w/o Align" achieves the worst performance due to the significant gap between LLM and recommendation domains. Among the alignment strategies, LEARN with ItemTower(a) achieves the best performance, followed by ItemTower(b), with ItemTower(c) performing the worst. We believe that LEARN with ItemTower(a), which uses sequence-to-sequence alignment, allows the model to better capture long-term user interests compared to the sequence-to-item alignment used in ItemTower(b). LEARN with ItemTower(c), which uses ``Rec-to-LLM" adaptation to align with the content embedding target, performs worse than ItemTower(a). This result is consistent with the findings on the Amazon Review datasets. 
Due to the lengthy user interactions spanning more than ten months, we apply the sample weighting strategy proposed in Eq.\ref{eq:weight} and compare it with a random sampling strategy. As shown in Tab.\ref{tab:ablation_indus}, LEARN with sample weighting achieves 7.13\% and 7.18\% improvement in H@100 and R@100.

\subsubsection{ID Embedding VS Content Embedding}
Given the limitations of ID embeddings in semantic representation and generalization, we explor the feasibility of alternatives to ID embeddings in large-scale real-world industrial scenarios. We used three approaches for item representation: learnable ID embeddings, frozen content embeddings extracted from the pretrained BERT \cite{chinese-clip} and pretrained LLM. The ID-emb dimension is set to 64 to align with the online system. As shown in Tab.~\ref{tab:inputemb}, LLM-based content embeddings deliver significant performance improvement over ID embeddings, with H@100 increasing from 0.0504 to 0.0751, representing 49.01\% enhancement. Compared to BERT-based content embeddings, LEARN with LLM embeddings achieves 30.38\% improvement. This can be attributed to the richer information contained in the LLM embeddings, which are trained on extensive text corpora. Our experimental results give a promising direction to replace ID embedding with semantic content embedding.

\begin{table}[t]
\centering
\resizebox{1\columnwidth}{!}{%
\begin{tabular}{l|c|cccc}
\hline
\textbf{Ablation} & \textbf{Params} & \textbf{H@50} & \textbf{R@50} & \textbf{H@100} & \textbf{R@100} \\ \hline
ID-emb & 2.3B & 0.0312 & 0.0499 & 0.0503 & 0.0754 \\
BERT-emb & 86M & 0.0357 & 0.0552 & 0.0576 & 0.0843 \\ 
LLM-emb (Ours)  & 89M & \textbf{0.0477} & \textbf{0.0663} & \textbf{0.0751} & \textbf{0.0970} \\ \hline
\end{tabular}%
}
\vspace{-1mm}
\caption{Performance comparison of input embedding types in the LEARN framework on the industry dataset. The number of trainable parameters is termed ``Params".}
\label{tab:inputemb}
\vspace{-3mm}
\end{table}

\subsubsection{Ablation studies of PAL Module}

\begin{table}[t]
\centering
\resizebox{\columnwidth}{!}{%
\begin{tabular}{c|cc|cc}
\toprule
 Ablation & Finetune & Params. &
H@100 & R@100 \\ \midrule
\multirow{3}{*}{w/ LLM} & \multirow{3}{*}{LoRA} & 134M & 0.0376 & 0.0560 \\
& & 286M & 0.0504 & 0.0709 \\
& & 572M & 0.0513 & 0.0720 \\\midrule
LEARN (Ours) &  Full & 89M & \textbf{0.0751} & \textbf{0.0970} \\ \bottomrule
\end{tabular}%
}
\vspace{-1mm}
\caption{Ablation studies of the backbone of PAL module. The ``w/ LLM" variant uses a pretrained LLM as the backbone of the PAL instead of 12 transformer layers. The number of trainable parameters is termed ``Params". }
\label{tab:LLM_lora}
\vspace{-4mm}
\end{table}

Given the superior capabilities of LLMs in text comprehension and common sense reasoning, we replace the transformer layers trained from scratch with the pretrained LLM (Baichuan2-7B) in the PAL module of Fig.~\ref{fig:tow_tower}. To retain open-world knowledge, we fine-tune the LLM using LoRA \cite{hu2021lora} and adjust different settings to vary the number of trainable parameters.

As shown in Tab.~\ref{tab:LLM_lora}, as the number of trainable parameters increased from 134M to 572M, the performance of variant ``w/ LLM" improves from 0.0376 to 0.0513. However, this performance still falls short compared to LEARN, which uses 12 transformer layers trained from scratch as the backbone. Considering the working mechanism of LoRA \cite{hu2021lora}, the output features of the ``w/ LLM" variant are a blend of the original feature trained in the open-world domain and the LoRA feature trained in the recommendation domain. 
Due to the significantly larger number of frozen parameters in the LLM compared to the trainable LoRA parameters, the original features tend to dominate. These original features are learned from the open world domain and are supervised by next-token prediction loss. In contrast, LoRA features are trained in the recommendation domain and are supervised by contrastive loss. This disparity between the two types of features prevents the mixed features from achieving optimal performance.

\subsection{Online A/B Experiments} \label{sec:online}

We evaluate the LEARN framework by online A/B testing on the ranking model of a popular short video streaming platform with more than 400 million daily active users (DAU). Our method has been deployed in the short video feed advertising scenario since January 2024. More details are provided in the supplementary.

\subsubsection{Ranking Model with LEARN Adaptor}
\begin{figure}[t]
    \centering
    \includegraphics[width=1\columnwidth]{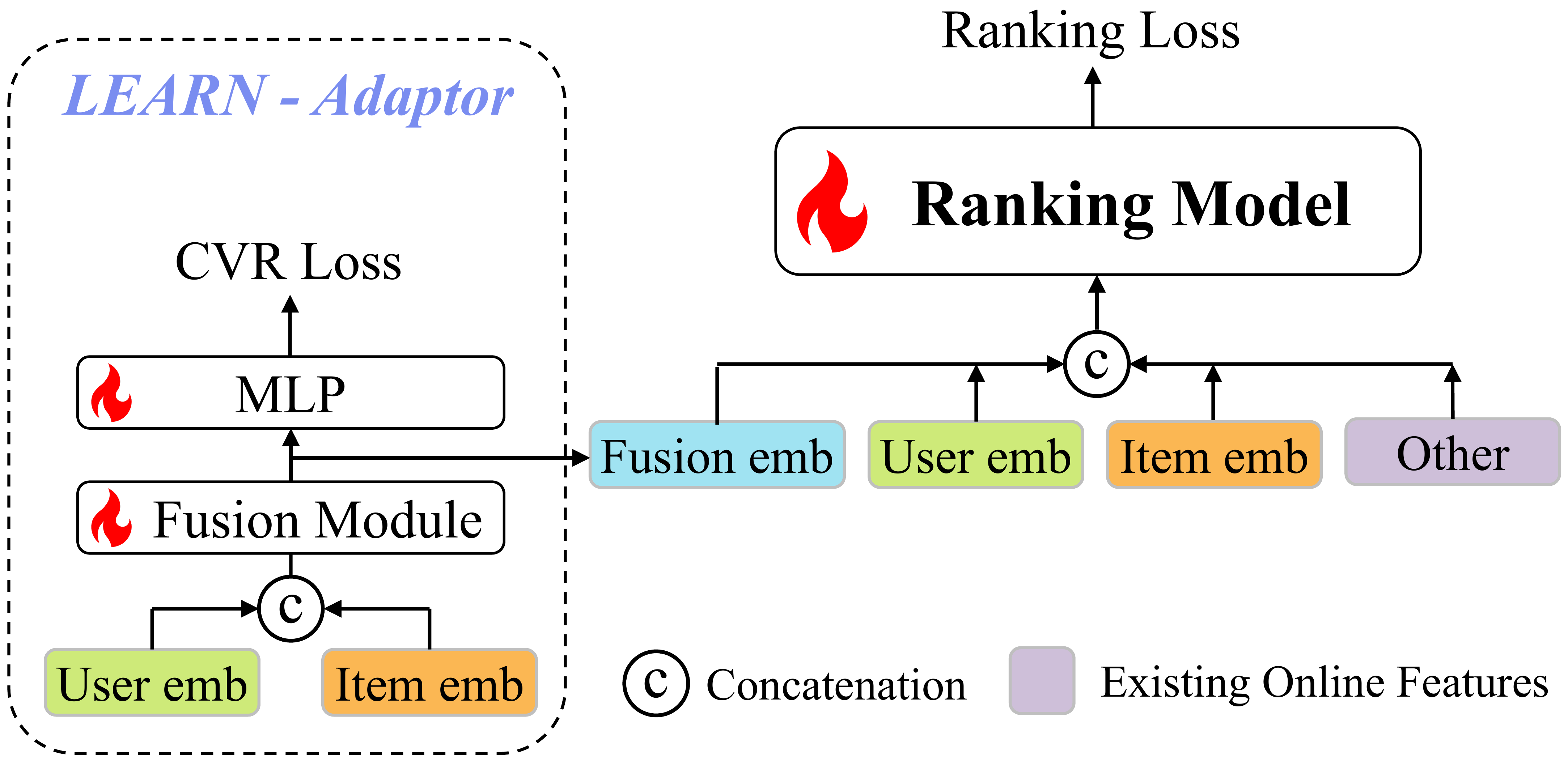}
    \vspace{-1em}
    \caption{Illustration of online ranking model structure.}
    \label{fig:cvr_model}
    \vspace{-1em}
\end{figure}

To better align the user and item embeddings generated by LEARN with the online ranking model, we introduced the \textit{LEARN-Aaptor} based on the baseline model. As illustrated in Fig.\ref{fig:cvr_model}, the baseline model consists of the original ranking model, which takes the existing online features as inputs (denoted as ``Other" in Fig.\ref{fig:cvr_model}). The \textit{LEARN-Adaptor} module includes a fusion module (two linear layers) and an MLP, which aggregate user and item embeddings into the fusion embedding through ConVersion Rate (CVR) loss. The fusion embedding, along with the user and item embedding of LEARN, and existing online features, are concatenated and fed into the ranking model.

\begin{table}[t]
\centering
\resizebox{1\columnwidth}{!}{%
\begin{tabular}{c|lll}
\hline
Method & UAUC & WUAUC \\ \hline
Baseline & 0.6885 & 0.7002 \\
LEARN (Ours) & 0.6969 (+0.84pp) & 0.7078 (+0.76pp) \\ \hline
\end{tabular}
}
\vspace{-1mm}
\caption{AUC results on our e-commerce platform.}
\label{tab:offline_auc}
\vspace{-6mm}
\end{table}

\begin{table}[t]
\centering
\resizebox{1\columnwidth}{!}{%
\begin{tabular}{c|l|rcc}
\hline
Level & Type & Proportion & Revenue & AUC \\ \hline
\multirow{3}{*}{User} & cold-start & 7.16\% & +1.56\% & +0.17pp \\
 & long-tail & 27.54\% & +5.79\% & +0.68pp \\ 
 & others & 65.30\% & +0.32\% & +0.021pp \\ \hline
\multirow{3}{*}{Item} & cold-start & 3.15\% &  +8.77\% & +0.29pp \\ 
& long-tail & 26.47\% &  +4.63\% & +0.21pp \\ 
& others & 70.38\% & +0.35\% & +0.01pp \\ \hline
\end{tabular}%
}
\vspace{-2mm}
\caption{Revenue improvement for three user and item types. ``Proportion" represents the percentage of users/items in each category.}
\label{tab:user_layers}
\vspace{-4mm}
\end{table}

\begin{figure}[t]
    \centering
    \includegraphics[width=0.9\columnwidth]{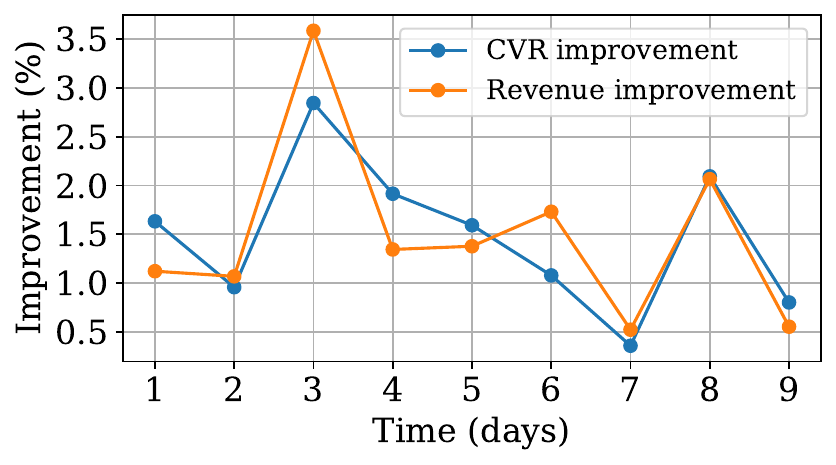}
    \vspace{-3mm}
    \caption{CVR and revenue improvements during online A/B testing. Compared to the baseline method, LEARN achieves a steady and significant increase. }
    \label{fig:Abtest}
    \vspace{-6mm}
\end{figure}

\subsubsection{AUC Evaluation}
AUC evaluation is conducted on a billion-scale dataset from the e-commerce platform of the short video application. Following common industry practices, we adopt UAUC and WUAUC metrics, as they more accurately assess the ranking performance for each user and better reflect user experience. Specifically, UAUC provides insight into the performance of long-tail users by applying uniform weights.
As depicted in Tab.~\ref{tab:offline_auc}, our method outperforms the baseline model, achieving improvements of 0.84 percentage points (pp) in UAUC and 0.76 pp in WUAUC. These gains can be attributed to the superior generalization capabilities of the LEARN framework, which effectively captures the interests of long-tail users.
To further validate this hypothesis, we conducted a more comprehensive analysis. We categorize users and items into three types based on historical interaction frequency. As shown in Tab.~\ref{tab:user_layers}, LEARN delivers significant performance enhancements, particularly for cold-start and long-tail users and items, further confirming the generalization of our approach for users and items with sparse purchase histories.

\subsubsection{Online Revenue Improvement}

We allocate 20\% of the platform's traffic to our proposed LEARN and baseline models. The experiments are conducted in a real-time system over 9 days. The results are shown in Fig.~\ref{fig:Abtest}. The LEARN framework demonstrates a steady and significant increase in both revenue and CVR. Considering that the revenue for online recommendation models is measured in tens of millions, even a 2\% improvement is highly significant.

\vspace{-2mm}
\section{Conclusion}

We explore integrating LLMs with recommendation systems and propose the LEARN framework to achieve significant business benefits.
The LEARN framework includes CEX and PAL modules. The CEX module uses the pretrained LLM to extract content embeddings for each item, while the PAL module projects these embeddings from the open-world domain to the user/item embeddings in the recommendation domain.
The state-of-the-art performance achieved in industry datasets and the public Amazon Review dataset demonstrates the superior performance of our LEARN framework.

\clearpage
\bibliography{main}

\begin{thebibliography}{44}
\providecommand{\natexlab}[1]{#1}

\bibitem[{Baichuan(2023)}]{baichuan2023baichuan2}
Baichuan. 2023.
\newblock Baichuan 2: Open Large-scale Language Models.
\newblock \emph{arXiv preprint arXiv:2309.10305}.

\bibitem[{Bao et~al.(2023{\natexlab{a}})Bao, Zhang, Wang, Zhang, Yang, Luo, Feng, He, and Tian}]{bao2023bi}
Bao, K.; Zhang, J.; Wang, W.; Zhang, Y.; Yang, Z.; Luo, Y.; Feng, F.; He, X.; and Tian, Q. 2023{\natexlab{a}}.
\newblock A bi-step grounding paradigm for large language models in recommendation systems.
\newblock \emph{arXiv preprint arXiv:2308.08434}.

\bibitem[{Bao et~al.(2023{\natexlab{b}})Bao, Zhang, Zhang, Wang, Feng, and He}]{bao2023tallrec}
Bao, K.; Zhang, J.; Zhang, Y.; Wang, W.; Feng, F.; and He, X. 2023{\natexlab{b}}.
\newblock Tallrec: An effective and efficient tuning framework to align large language model with recommendation.
\newblock In \emph{Proceedings of the 17th ACM Conference on Recommender Systems}, 1007--1014.

\bibitem[{Chen et~al.(2023)Chen, Li, Dong, Zhang, Wang, and Wang}]{chen2023tbin}
Chen, S.; Li, X.; Dong, J.; Zhang, J.; Wang, Y.; and Wang, X. 2023.
\newblock TBIN: Modeling Long Textual Behavior Data for CTR Prediction.
\newblock \emph{arXiv preprint arXiv:2308.08483}.

\bibitem[{Chiang et~al.(2023)Chiang, Li, Lin, Sheng, Wu, Zhang, Zheng, Zhuang, Zhuang, Gonzalez et~al.}]{chiang2023vicuna}
Chiang, W.-L.; Li, Z.; Lin, Z.; Sheng, Y.; Wu, Z.; Zhang, H.; Zheng, L.; Zhuang, S.; Zhuang, Y.; Gonzalez, J.~E.; et~al. 2023.
\newblock Vicuna: An open-source chatbot impressing gpt-4 with 90\%* chatgpt quality.
\newblock \emph{See https://vicuna. lmsys. org (accessed 14 April 2023)}, 2(3): 6.

\bibitem[{Devlin et~al.(2018)Devlin, Chang, Lee, and Toutanova}]{devlin2018bert}
Devlin, J.; Chang, M.-W.; Lee, K.; and Toutanova, K. 2018.
\newblock Bert: Pre-training of deep bidirectional transformers for language understanding.
\newblock \emph{arXiv preprint arXiv:1810.04805}.

\bibitem[{Ding et~al.(2021)Ding, Ma, Deoras, Wang, and Wang}]{ding2021zero}
Ding, H.; Ma, Y.; Deoras, A.; Wang, Y.; and Wang, H. 2021.
\newblock Zero-shot recommender systems.
\newblock \emph{arXiv preprint arXiv:2105.08318}.

\bibitem[{Dosovitskiy et~al.(2020)Dosovitskiy, Beyer, Kolesnikov, Weissenborn, Zhai, Unterthiner, Dehghani, Minderer, Heigold, Gelly et~al.}]{dosovitskiy2020image}
Dosovitskiy, A.; Beyer, L.; Kolesnikov, A.; Weissenborn, D.; Zhai, X.; Unterthiner, T.; Dehghani, M.; Minderer, M.; Heigold, G.; Gelly, S.; et~al. 2020.
\newblock An image is worth 16x16 words: Transformers for image recognition at scale.
\newblock \emph{arXiv preprint arXiv:2010.11929}.

\bibitem[{Fan et~al.(2023)Fan, Zhao, Li, Liu, Mei, Wang, Tang, and Li}]{fan2023recommender}
Fan, W.; Zhao, Z.; Li, J.; Liu, Y.; Mei, X.; Wang, Y.; Tang, J.; and Li, Q. 2023.
\newblock Recommender systems in the era of large language models (llms).
\newblock \emph{arXiv preprint arXiv:2307.02046}.

\bibitem[{Gao et~al.(2023)Gao, Sheng, Xiang, Xiong, Wang, and Zhang}]{gao2023chat}
Gao, Y.; Sheng, T.; Xiang, Y.; Xiong, Y.; Wang, H.; and Zhang, J. 2023.
\newblock Chat-rec: Towards interactive and explainable llms-augmented recommender system.
\newblock \emph{arXiv preprint arXiv:2303.14524}.

\bibitem[{He et~al.(2020)He, Fan, Wu, Xie, and Girshick}]{he2020momentum}
He, K.; Fan, H.; Wu, Y.; Xie, S.; and Girshick, R. 2020.
\newblock Momentum contrast for unsupervised visual representation learning.
\newblock In \emph{Proceedings of the IEEE/CVF conference on computer vision and pattern recognition}, 9729--9738.

\bibitem[{He et~al.(2016)He, Zhang, Ren, and Sun}]{he2016deep}
He, K.; Zhang, X.; Ren, S.; and Sun, J. 2016.
\newblock Deep residual learning for image recognition.
\newblock In \emph{Proceedings of the IEEE conference on computer vision and pattern recognition}, 770--778.

\bibitem[{Hidasi et~al.(2015)Hidasi, Karatzoglou, Baltrunas, and Tikk}]{hidasi2015session}
Hidasi, B.; Karatzoglou, A.; Baltrunas, L.; and Tikk, D. 2015.
\newblock Session-based recommendations with recurrent neural networks.
\newblock \emph{arXiv preprint arXiv:1511.06939}.

\bibitem[{Hou et~al.(2022)Hou, Mu, Zhao, Li, Ding, and Wen}]{hou2022towards}
Hou, Y.; Mu, S.; Zhao, W.~X.; Li, Y.; Ding, B.; and Wen, J.-R. 2022.
\newblock Towards universal sequence representation learning for recommender systems.
\newblock In \emph{Proceedings of the 28th ACM SIGKDD Conference on Knowledge Discovery and Data Mining}, 585--593.

\bibitem[{Hou et~al.(2024)Hou, Zhang, Lin, Lu, Xie, McAuley, and Zhao}]{hou2024large}
Hou, Y.; Zhang, J.; Lin, Z.; Lu, H.; Xie, R.; McAuley, J.; and Zhao, W.~X. 2024.
\newblock Large language models are zero-shot rankers for recommender systems.
\newblock In \emph{European Conference on Information Retrieval}, 364--381. Springer.

\bibitem[{Hu et~al.(2021)Hu, Shen, Wallis, Allen-Zhu, Li, Wang, Wang, and Chen}]{hu2021lora}
Hu, E.~J.; Shen, Y.; Wallis, P.; Allen-Zhu, Z.; Li, Y.; Wang, S.; Wang, L.; and Chen, W. 2021.
\newblock Lora: Low-rank adaptation of large language models.
\newblock \emph{arXiv preprint arXiv:2106.09685}.

\bibitem[{Kang and McAuley(2018)}]{kang2018self}
Kang, W.-C.; and McAuley, J. 2018.
\newblock Self-attentive sequential recommendation.
\newblock In \emph{2018 IEEE international conference on data mining (ICDM)}, 197--206. IEEE.

\bibitem[{Li et~al.(2022)Li, Zhao, Zhang, Yu, Cheng, Shu, Kong, and Niu}]{li2022recguru}
Li, C.; Zhao, M.; Zhang, H.; Yu, C.; Cheng, L.; Shu, G.; Kong, B.; and Niu, D. 2022.
\newblock RecGURU: Adversarial learning of generalized user representations for cross-domain recommendation.
\newblock In \emph{Proceedings of the fifteenth ACM international conference on web search and data mining}, 571--581.

\bibitem[{Li et~al.(2023)Li, Wang, Li, Fu, Shen, Shang, and McAuley}]{li2023text}
Li, J.; Wang, M.; Li, J.; Fu, J.; Shen, X.; Shang, J.; and McAuley, J. 2023.
\newblock Text is all you need: Learning language representations for sequential recommendation.
\newblock In \emph{Proceedings of the 29th ACM SIGKDD Conference on Knowledge Discovery and Data Mining}, 1258--1267.

\bibitem[{Li, Zhang, and Chen(2023)}]{li2023prompt}
Li, L.; Zhang, Y.; and Chen, L. 2023.
\newblock Prompt distillation for efficient llm-based recommendation.
\newblock In \emph{Proceedings of the 32nd ACM International Conference on Information and Knowledge Management}, 1348--1357.

\bibitem[{Liao et~al.(2023)Liao, Li, Yang, Wu, Yuan, Wang, and He}]{liao2023llara}
Liao, J.; Li, S.; Yang, Z.; Wu, J.; Yuan, Y.; Wang, X.; and He, X. 2023.
\newblock Llara: Aligning large language models with sequential recommenders.
\newblock \emph{arXiv preprint arXiv:2312.02445}.

\bibitem[{Lin et~al.(2023{\natexlab{a}})Lin, Dai, Xi, Liu, Chen, Li, Zhu, Guo, Yu, Tang et~al.}]{lin2023can}
Lin, J.; Dai, X.; Xi, Y.; Liu, W.; Chen, B.; Li, X.; Zhu, C.; Guo, H.; Yu, Y.; Tang, R.; et~al. 2023{\natexlab{a}}.
\newblock How can recommender systems benefit from large language models: A survey.
\newblock \emph{arXiv preprint arXiv:2306.05817}.

\bibitem[{Lin et~al.(2023{\natexlab{b}})Lin, Shan, Zhu, Du, Chen, Quan, Tang, Yu, and Zhang}]{lin2023rella}
Lin, J.; Shan, R.; Zhu, C.; Du, K.; Chen, B.; Quan, S.; Tang, R.; Yu, Y.; and Zhang, W. 2023{\natexlab{b}}.
\newblock Rella: Retrieval-enhanced large language models for lifelong sequential behavior comprehension in recommendation.
\newblock \emph{arXiv preprint arXiv:2308.11131}.

\bibitem[{Liu et~al.(2023)Liu, Liu, Lv, Zhou, and Zhang}]{liu2023chatgpt}
Liu, J.; Liu, C.; Lv, R.; Zhou, K.; and Zhang, Y. 2023.
\newblock Is chatgpt a good recommender? a preliminary study.
\newblock \emph{arXiv preprint arXiv:2304.10149}.

\bibitem[{Lyu et~al.(2023)Lyu, Jiang, Zeng, Xia, and Luo}]{lyu2023llm}
Lyu, H.; Jiang, S.; Zeng, H.; Xia, Y.; and Luo, J. 2023.
\newblock Llm-rec: Personalized recommendation via prompting large language models.
\newblock \emph{arXiv preprint arXiv:2307.15780}.

\bibitem[{Ni, Li, and McAuley(2019)}]{ni2019justifying}
Ni, J.; Li, J.; and McAuley, J. 2019.
\newblock Justifying recommendations using distantly-labeled reviews and fine-grained aspects.
\newblock In \emph{Proceedings of the 2019 conference on empirical methods in natural language processing and the 9th international joint conference on natural language processing (EMNLP-IJCNLP)}, 188--197.

\bibitem[{Pancha et~al.(2022)Pancha, Zhai, Leskovec, and Rosenberg}]{pancha2022pinnerformer}
Pancha, N.; Zhai, A.; Leskovec, J.; and Rosenberg, C. 2022.
\newblock PinnerFormer: Sequence Modeling for User Representation at Pinterest.
\newblock In \emph{Proceedings of the 28th ACM SIGKDD Conference on Knowledge Discovery and Data Mining}, 3702--3712.

\bibitem[{Radford et~al.(2021)Radford, Kim, Hallacy, Ramesh, Goh, Agarwal, Sastry, Askell, Mishkin, Clark et~al.}]{radford2021learning}
Radford, A.; Kim, J.~W.; Hallacy, C.; Ramesh, A.; Goh, G.; Agarwal, S.; Sastry, G.; Askell, A.; Mishkin, P.; Clark, J.; et~al. 2021.
\newblock Learning transferable visual models from natural language supervision.
\newblock In \emph{International conference on machine learning}, 8748--8763. PMLR.

\bibitem[{Ren et~al.(2023)Ren, Wei, Xia, Su, Cheng, Wang, Yin, and Huang}]{ren2023representation}
Ren, X.; Wei, W.; Xia, L.; Su, L.; Cheng, S.; Wang, J.; Yin, D.; and Huang, C. 2023.
\newblock Representation learning with large language models for recommendation.
\newblock \emph{arXiv preprint arXiv:2310.15950}.

\bibitem[{Ren et~al.(2024)Ren, Wei, Xia, Su, Cheng, Wang, Yin, and Huang}]{ren2024representation}
Ren, X.; Wei, W.; Xia, L.; Su, L.; Cheng, S.; Wang, J.; Yin, D.; and Huang, C. 2024.
\newblock Representation learning with large language models for recommendation.
\newblock In \emph{Proceedings of the ACM on Web Conference 2024}, 3464--3475.

\bibitem[{Sanner et~al.(2023)Sanner, Balog, Radlinski, Wedin, and Dixon}]{zs-llm4rec-sanner2023large}
Sanner, S.; Balog, K.; Radlinski, F.; Wedin, B.; and Dixon, L. 2023.
\newblock Large language models are competitive near cold-start recommenders for language-and item-based preferences.
\newblock In \emph{Proceedings of the 17th ACM conference on recommender systems}, 890--896.

\bibitem[{Sheng et~al.(2024)Sheng, Yang, Gong, Wang, Chan, Zhang, Cheng, Zhu, Ge, Zhu et~al.}]{sheng2024enhancing}
Sheng, X.-R.; Yang, F.; Gong, L.; Wang, B.; Chan, Z.; Zhang, Y.; Cheng, Y.; Zhu, Y.-N.; Ge, T.; Zhu, H.; et~al. 2024.
\newblock Enhancing Taobao Display Advertising with Multimodal Representations: Challenges, Approaches and Insights.
\newblock \emph{arXiv preprint arXiv:2407.19467}.

\bibitem[{Sun et~al.(2019)Sun, Liu, Wu, Pei, Lin, Ou, and Jiang}]{sun2019bert4rec}
Sun, F.; Liu, J.; Wu, J.; Pei, C.; Lin, X.; Ou, W.; and Jiang, P. 2019.
\newblock BERT4Rec: Sequential recommendation with bidirectional encoder representations from transformer.
\newblock In \emph{Proceedings of the 28th ACM international conference on information and knowledge management}, 1441--1450.

\bibitem[{Wang et~al.(2023)Wang, Zeng, Wang, Wang, Lu, Li, Yuan, Zhang, Zheng, and Xia}]{wang2023missrec}
Wang, J.; Zeng, Z.; Wang, Y.; Wang, Y.; Lu, X.; Li, T.; Yuan, J.; Zhang, R.; Zheng, H.-T.; and Xia, S.-T. 2023.
\newblock Missrec: Pre-training and transferring multi-modal interest-aware sequence representation for recommendation.
\newblock In \emph{Proceedings of the 31st ACM International Conference on Multimedia}, 6548--6557.

\bibitem[{Wu et~al.(2020)Wu, Qiao, Chen, Wu, Qi, Lian, Liu, Xie, Gao, Wu et~al.}]{wu2020mind}
Wu, F.; Qiao, Y.; Chen, J.-H.; Wu, C.; Qi, T.; Lian, J.; Liu, D.; Xie, X.; Gao, J.; Wu, W.; et~al. 2020.
\newblock Mind: A large-scale dataset for news recommendation.
\newblock In \emph{Proceedings of the 58th annual meeting of the association for computational linguistics}, 3597--3606.

\bibitem[{Xi et~al.(2023)Xi, Liu, Lin, Zhu, Chen, Tang, Zhang, Zhang, and Yu}]{xi2023towards}
Xi, Y.; Liu, W.; Lin, J.; Zhu, J.; Chen, B.; Tang, R.; Zhang, W.; Zhang, R.; and Yu, Y. 2023.
\newblock Towards open-world recommendation with knowledge augmentation from large language models.
\newblock \emph{arXiv preprint arXiv:2306.10933}.

\bibitem[{Yang et~al.(2022)Yang, Pan, Lin, Men, Zhang, Zhou, and Zhou}]{chinese-clip}
Yang, A.; Pan, J.; Lin, J.; Men, R.; Zhang, Y.; Zhou, J.; and Zhou, C. 2022.
\newblock Chinese CLIP: Contrastive Vision-Language Pretraining in Chinese.
\newblock \emph{arXiv preprint arXiv:2211.01335}.

\bibitem[{Yuan et~al.(2023)Yuan, Yuan, Song, Li, Fu, Yang, Pan, and Ni}]{yuan2023go}
Yuan, Z.; Yuan, F.; Song, Y.; Li, Y.; Fu, J.; Yang, F.; Pan, Y.; and Ni, Y. 2023.
\newblock Where to go next for recommender systems? id-vs. modality-based recommender models revisited.
\newblock In \emph{Proceedings of the 46th International ACM SIGIR Conference on Research and Development in Information Retrieval}, 2639--2649.

\bibitem[{Yue et~al.(2023)Yue, Rabhi, Moreira, Wang, and Oldridge}]{yue2023llamarec}
Yue, Z.; Rabhi, S.; Moreira, G. d. S.~P.; Wang, D.; and Oldridge, E. 2023.
\newblock LlamaRec: Two-stage recommendation using large language models for ranking.
\newblock \emph{arXiv preprint arXiv:2311.02089}.

\bibitem[{Zhang et~al.(2023{\natexlab{a}})Zhang, Bao, Zhang, Wang, Feng, and He}]{zhang2023chatgpt}
Zhang, J.; Bao, K.; Zhang, Y.; Wang, W.; Feng, F.; and He, X. 2023{\natexlab{a}}.
\newblock Is chatgpt fair for recommendation? evaluating fairness in large language model recommendation.
\newblock In \emph{Proceedings of the 17th ACM Conference on Recommender Systems}, 993--999.

\bibitem[{Zhang et~al.(2023{\natexlab{b}})Zhang, Xie, Hou, Zhao, Lin, and Wen}]{zhang2023recommendation}
Zhang, J.; Xie, R.; Hou, Y.; Zhao, W.~X.; Lin, L.; and Wen, J.-R. 2023{\natexlab{b}}.
\newblock Recommendation as instruction following: A large language model empowered recommendation approach.
\newblock \emph{arXiv preprint arXiv:2305.07001}.

\bibitem[{Zhang et~al.(2019)Zhang, Zhao, Liu, Sheng, Xu, Wang, Liu, Zhou et~al.}]{zhang2019feature}
Zhang, T.; Zhao, P.; Liu, Y.; Sheng, V.~S.; Xu, J.; Wang, D.; Liu, G.; Zhou, X.; et~al. 2019.
\newblock Feature-level deeper self-attention network for sequential recommendation.
\newblock In \emph{IJCAI}, 4320--4326.

\bibitem[{Zhang et~al.(2024)Zhang, Yu, Zhang, Chen, Hu, Jiang, and Gai}]{zhang2024recgpt}
Zhang, Y.; Yu, W.; Zhang, E.; Chen, X.; Hu, L.; Jiang, P.; and Gai, K. 2024.
\newblock RecGPT: Generative Personalized Prompts for Sequential Recommendation via ChatGPT Training Paradigm.
\newblock \emph{arXiv preprint arXiv:2404.08675}.

\bibitem[{Zhou et~al.(2020)Zhou, Wang, Zhao, Zhu, Wang, Zhang, Wang, and Wen}]{zhou2020s3}
Zhou, K.; Wang, H.; Zhao, W.~X.; Zhu, Y.; Wang, S.; Zhang, F.; Wang, Z.; and Wen, J.-R. 2020.
\newblock S3-rec: Self-supervised learning for sequential recommendation with mutual information maximization.
\newblock In \emph{Proceedings of the 29th ACM international conference on information \& knowledge management}, 1893--1902.

\end{thebibliography}
\clearpage

\end{document}